\documentstyle[11pt,aaspp4,psfig]{article}

\received{}
\accepted{}
\journalid{}{}
\articleid{}{}

\slugcomment{ to appear in the Astrophysical Journal}

\lefthead{Krennrich et al.}
\righthead{TeV Flare-S}

\begin{document}

\def\lsim{\mathrel{\hbox{\rlap{\hbox{\lower4pt\hbox{$\sim$}}}\hbox{$<$}}}}
\def\gsim{\mathrel{\hbox{\rlap{\hbox{\lower4pt\hbox{$\sim$}}}\hbox{$>$}}}}

\title{Detection Techniques of Microsecond Gamma-Ray Bursts using Ground-Based Telescopes  }

\author{F.~Krennrich\altaffilmark{1},  S. Le Bohec\altaffilmark{1} \& T.C. Weekes\altaffilmark{2}}

\altaffiltext{1}{Department of Physics and Astronomy, Iowa State
University, Ames, IA 50011-3160}

\altaffiltext{2}{ Fred Lawrence Whipple Observatory, Harvard-Smithsonian 
CfA, P.O.~Box 97, Amado, AZ 85645-0097}

\clearpage
\begin{abstract}

Gamma-ray observations above 200~MeV are conventionally made
 by satellite-based detectors. The  EGRET detector on
the Compton Gamma Ray Observatory (CGRO) has  provided good  sensitivity for 
the detection of bursts lasting  for more than 200~ms. 
Theoretical predictions of high-energy $\gamma$-ray bursts produced
by quantum-mechanical decay of primordial black holes  (Hawking 1971)
suggest the emission of bursts on  shorter time scales. 
The final stage of a primordial black hole results in a burst 
of $\gamma$-rays,  peaking around 250 MeV and lasting for a tenth of a microsecond 
or longer depending on particle physics.
In this work we show that there is an observational window using
ground-based imaging Cherenkov  detectors  to measure  $\gamma$-ray burst 
emission at energies E~$\rm >$~200~MeV.    This technique,  with a sensitivity 
for bursts lasting nanoseconds to several microseconds, is based on the detection 
of multi-photon-initiated air showers.

\end{abstract}


\section{Introduction}

The astrophysical band for the detection of high energy $\gamma$-rays has been 
recently expanded to energies between hundreds of GeV (Weekes et al. 1989) up 
to beyond 10~TeV (Aharonian et al. 1997; Tanimori et al. 1998a; Krennrich et al. 1999a) 
using the ground-based atmospheric Cherenkov imaging technique.   The proposed
coverage from 20~MeV~-~300~GeV with the future satellite-based GLAST detector 
(Gehrels \& Michelson 1999) providing a large field of view is complemented by 
the proposals of ground-based detectors such as VERITAS (Weekes et al. 1999), HESS 
(Hofmann et al. 1997)  and MAGIC (Barrio et al. 1998) with an energy threshold
in the tens of GeV range.   Ground-based  Cherenkov imaging detectors 
provide large collection areas of the order of $\rm 10^5~m^2$ and hence, are  
well suited to the study of $\gamma$-ray flare    phenomena.    This technique has 
already proven successful in the study of  AGN flares on minute time scales 
(Gaidos et al. 1996) and is expected to improve in sensitivity  by an order 
of magnitude with future detectors.

In this paper we explore the possibility of using imaging atmospheric Cherenkov 
telescopes to detect $\gamma$-ray flare phenomena on shorter time scales
of microseconds and with energies in the sub-GeV regime.  
     Astrophysical phenomena producing extremely short bursts of $\gamma$-rays could 
be the  signature of Hawking's prediction of   $\gamma$-ray burst radiation from 
the evaporation of primordial black holes  (Hawking 1974).  The lifetime of a black hole
is proportional to the mass cubed.            In the early Universe 
primordial black holes (PBHs) of small mass  may have formed (Hawking 1971; Carr 1976).   
PBHs created with initial masses  slightly greater than   
$\rm \approx 5 \times 10^{14} grams$ would 
 be evaporating now by the quantum-gravitational Hawking mechanism.   A PBH's 
existence ends in a dramatic explosion where the final stage of 
the evaporation is determined by particle physics at extremely high energies.
Hagedorn (1970) suggested a particle physics model in which  the number of 
species of particles increases exponentially with energy. In this scenario, 
a black hole loses its energy quickly when reaching a critical temperature.  
 A burst of $\gamma$-rays as short as $\rm 10^{-7} s$, with a total energy of 
$\rm 10^{34}$ ergs, would be the signature of such an event.   However, the burst 
time scale and average photon energy depends  on the particle physics model, 
with highly uncertain predictions at high energies. 
This has prompted searches over much larger time and energy  scales ranging from 
$\rm 10^{-7} s$ at 250~MeV to seconds at 10~TeV, as suggested by the standard model of 
particle physics (Halzen et al. 1991).   Cline \& Hong (1992), using a mixture of a Hagedorn
and QCD-like spectrum,  suggested that these  bursts occur on the millisecond time scale 
in the MeV range.  

Classical $\gamma$-ray bursts (GRBs) detected with the satellite-experiment BATSE 
on CGRO show $\gamma$-ray emission on surprisingly short time 
scales. GRB time scales
in the millisecond range have been reported by Kouveliotou et al. (1994) -  
the detection of the so-called ``Superbowl'' burst (GRB930131) has revealed 
temporal variations on time scales   as short as 2 ms.  In fact, evidence for 
sub-millisecond  (200 $\rm \mu sec$) structures was found in the BATSE data of 
GRB910305 (Bhat et al. 1992).     EGRET, which was sensitive from 30~MeV to 30~GeV,
due to an instrumental dead time effect,  was limited in sensitivity for short 
bursts to time  scales above 200~ms.

A search for microsecond scale  bursts using EGRET has been  made by looking for 
multiple-$\gamma$-ray events arriving almost   simultaneously (within a single
spark chamber gate, i.e., 600 ns); it  produced only an upper limit of 
$\rm 5 \times 10^{-2} /yr/pc^3$  (Fichtel et al. 1994). 
Searches by Cline et al. (1997) using archival data from the BATSE experiment found 
some events on millisecond time scales, but it was not possible to prove that they were
not just classical $\gamma$-ray bursters.    Also in an early experiment, 
first generation ground-based atmospheric Cherenkov detectors were used to search 
on the shortest time-scales predicted ($\rm 10^{-7} s$), giving an upper limit of 
$\rm 4 \times 10^{-2} /yr/pc^3$ (Porter \& Weekes 1978).    The possibility of using 
atmospheric Cherenkov imaging telescopes to detect wavefront events was considered 
elsewhere (Connaughton, 1996); for a single telescope with a camera with relatively small
field of view it was shown to be difficult to recognize the bursts
and distinguish them from background cosmic-ray events.

 The technique described here can be used to search for microsecond $\gamma$-ray 
emission in the sub-GeV regime with a more sensitive ground-based instrument.
The proposed detection technique will clearly identify these events.
A short  burst can be approximated as a thin  plane wavefront 
of $\gamma$-rays traveling through space (wavefront event, hereafter),
starting a multi-photon-initiated cascade when entering the earth's atmosphere. 
Measuring the angular distribution of Cherenkov light from a short burst 
using an atmospheric Cherenkov imaging detector is a new approach  to distinguish 
short  bursts from background by cosmic rays.  Previous efforts (Porter \& Weekes 1978)
used non-imaging Cherenkov detectors, and the suppression of cosmic rays  was achieved by 
simultaneous recording  by two telescopes separated at a distance of 400~km.   
Imaging  enables  the identification of a $\gamma$-ray  wavefront event in a
single telescope  and the  measurement of its arrival direction.    With some modifications (\S 4)
 future ground-based $\gamma$-ray detectors using  arrays of imaging telescopes, e.g., 
VERITAS (Weekes et al. 1999) and HESS (Hofmann et al. 1997),  would be ideally suited  
for exploring this  observational window of  microsecond  bursts.

In \S 2 we describe the phenomenology of the wavefront events and how they
differ from single-particle-initiated air showers. 
In \S 3, using Monte Carlo simulations,  we describe an analysis technique including 
timing characteristics to separate wavefront events from background arising from cosmic-ray 
showers.   We also discuss the design  considerations (\S 4) for the implementation in 
imaging Cherenkov telescopes.  In \S 5, an estimate of flux sensitivity and the 
energy range of the existing Whipple Observatory 10~m  instrument are shown.

\section{Phenomenology of multi-photon-initiated showers}

The technique proposed here builds upon the atmospheric Cherenkov imaging 
technique that has been pivotal in establishing the field of TeV $\gamma$-ray 
astrophysics (for review see Ong et al. 1998).  The  technique provides
the highest sensitivity for  detecting  $\gamma$-ray sources above 200~GeV.
In this technique, Cherenkov light from an electromagnetic atmospheric cascade is focused 
onto a camera of fast photomultiplier tubes.  The  images are analyzed to select 
$\gamma$-ray events while rejecting over   99.7\% of cosmic-ray background events.   
This has led to the discovery of TeV   $\gamma$-rays from the Crab Nebula (Weekes et al. 1989),
PSR~1706-44 (Kifune et al. 1995), Vela  (Yoshikoshi et al. 1997) and SN~1006 
(Tanimori et al. 1998b) and 
from three active galactic  nuclei: Mrk~421 (Punch et al. 1992),  Mrk 501 (Quinn et al. 1996)
and 1ES~2344+514 (Catanese et al. 1998).
Atmospheric Cherenkov telescopes have a high collection area ($\rm \sim 50,000 \: m^2$ for 
a  single 250 GeV $\gamma$-ray) making them uniquely sensitive  to short time-scale phenomena.  

Cherenkov light from a plane wavefront of multiple $\rm E \: >$~200~MeV $\gamma$-rays 
 can be detected with ground-based optical telescopes.  
A low energy multi-photon-initiated cascade differs significantly from a 
single-particle-initiated cascade e.g., a TeV photon or proton induced  shower.    
Individual low energy $\gamma$-rays, when reaching  the upper atmosphere, will typically 
generate one or a few (depending on energy) generations of electrons and positrons 
(collectively called electrons hereafter) by pair production and subsequent bremsstrahlung.
The electrons, before falling below the critical energy,   radiate Cherenkov light 
(6000 photons per electron for one radiation length)  
which can be collected by an optical reflector at ground level.  
The average  number of    Cherenkov    photons associated with a single sub-GeV 
$\gamma$-ray is small, and therefore, its  Cherenkov flash is too faint to be 
detectable at ground level.   However,   a large number of $\gamma$-rays arriving 
within a short time can  produce a Cherenkov signal strong enough to be detectable
by an atmospheric Cherenkov telescope.

Previous efforts to detect  wavefront events were based on the fact  that 
multi-photon-initiated   showers have a large lateral extent.   They can be detected by 
using  relatively simple non-imaging atmospheric Cherenkov  telescopes  (Porter \& Weekes 1978).   
For  logistical and cost  reasons it is difficult to operate two
telescopes at a distance several hundred miles apart, solely dedicated to a search for bursts.
On the contrary,  existing  imaging telescopes or future arrays of imaging 
telescopes,  can be used in parallel with standard TeV $\gamma$-ray observations to search 
for wavefront events from microsecond bursts.   These    instruments also provide a 
significant improvement  to previous efforts: the imaging 
capability provides clear recognition of the wavefront events  from the measurement
of  the angular Cherenkov light distribution in the focal plane combined with 
the Cherenkov pulse width.

There are three unique characteristics of the Cherenkov light image
produced by a wavefront event. 

a)  The first  is the very large  extent  of the wavefront, which means it can be detected 
simultaneously by telescopes over vast distances.  The images in all telescopes in an
array (for example the VERITAS array) should be identical, regardless of their distance.   
This is different  from single-particle-initiated shower images  which are detectable 
over a limited  area on the  ground and,  if detected, show a parallactic displacement
between telescopes.  

b) The second characteristic is the time profile of the Cherenkov pulse 
which can range from  $\rm \approx$~100~nanoseconds  to microseconds - and thus is quite 
different from  Cherenkov flashes of conventional air showers showing durations of
 5-30~nanosecond.   The Cherenkov light time profile of relatively 
 long (microsecond) bursts is dominated  by the intrinsic width of the burst itself. 
However, the time profile of a wavefront event is also determined by the geometry of the
multi-photon-initiated cascade:  the detection is based on collecting 
Cherenkov photons that have been emitted by secondary electrons of the cascades
initiated by $\gamma$-ray primaries with a large range of impact points.
Cherenkov photons can be collected  up to several hundred meters 
distance from the extrapolated impact point of the primary  at  ground-level.  
The intrinsic differences in time-of-flight between Cherenkov
photons from different primary particle impact points causes multi-photon-initiated
showers to have a minimum width of $\rm \approx 40$~nanoseconds 
(see \S 3.3), assuming the time profile of the $\gamma$-rays is a delta function. 
The time structure of the images shows a concentric symmetry: the closer to the
center, the earlier the pulse. 

c) The third characteristic,  which is hinted at by Figure~1, but is not entirely
obvious, is that the images in the camera plane from a wavefront event are circular.  They 
also will provide information about the arrival direction of the wavefront: 
the displacement of the image centroid from the optic axis of the telescope 
measures the arrival direction of the burst.

Figure 2 shows the simulated image (see \S 3.1) of a  Cherenkov flash from a 300~MeV 
$\gamma$-ray burst (pulse width of 100 ns with 0.5 $\rm \gamma's/m^2$;   fluence =
$\rm 2.4  \times  10^{-8} ergs/cm^2$) in the focal plane  of the Whipple 
Observatory 10~m telescope.  
The background from night-sky fluctuations for a 100~ns exposure has been included and 
a standard image cleaning procedure (Reynolds et al. 1993) applied.  The light distribution 
in the image center is relatively flat and smooth.  
The flatness of the light distribution arises from a uniform lateral density 
distribution of  electrons.   Shower fluctuations have very 
little effect on the Cherenkov light distribution 
because of the huge number of showers contributing to the Cherenkov flash.
This results in a smooth light distribution with mainly statistical 
variations due to the night sky background and instrumental noise. 

The Cherenkov light angular distribution is determined by the Cherenkov angle 
at a given height ($\rm 0.4^{\circ}$ at 15~km height) and  the multiple-scattering angle
of the electrons in the cascade.      The convolution of both effects leads 
to images that show a prominent plateau with  a radial extension of $\rm \approx 0.3^{\circ}$ 
with a ``halo'' extending further with a scale of $\rm \approx 2^{\circ}$ F.W.H.M. 
  These image shapes  clearly differ from single $\gamma$-ray or cosmic-ray initiated
Cherenkov images (Hillas 1996) and provide an important constraint for classifying
these short bursts. 
    Together, with the timing information of the Cherenkov pulse-shape,  imaging can 
be used to reject background events from cosmic rays.

Because we are describing a  burst detection technique,  fluence and sensitive area are used 
to describe the detector properties and are defined as follows:  

1.~Fluence: A detector is triggered whenever the number of $\gamma$-rays during an
integration time bin of some duration exceeds a threshold.  In the following, we 
 use the term fluence, the total energy S received from a given burst 
in units of $\rm ergs/ cm^{2}$  over the full duration of the burst.
Since we are not trying to resolve individual photons during the burst such 
an integral measure is sufficient.  

2. Sensitive area:  Cherenkov photons emitted by an electron at
 20~km  atmospheric height are most likely spread over an area of 500~m  in 
radius.   However, a few photons,  emitted from   electrons   with large 
multiple-scattering angles, reach up to 800~m   from the impact point of the primary 
$\gamma$-ray.   This results in a large sensitive area ($\rm 2 \times 10^6 \: m^2 $) 
over which  individual $\gamma$-rays make a contribution to the total amount of
light of a Cherenkov flash. 
The efficiency for a single sub-GeV $\gamma$-ray triggering a reasonable
sized  atmospheric Cherenkov imaging telescope ($\rm < \: 20$~m reflector diameter) 
is essentially zero.     For a 300~MeV~$\gamma$-ray,   the efficiency for contributing a 
single photoelectron in the photomultiplier camera of the Whipple Observatory 10~m telescope reaches a 
maximum of approximately 1\%.  
The sensitive area is the area for which an individual low energy shower makes a 
significant contribution to the Cherenkov light flash.

\section{Simulations}

We have carried out Monte Carlo simulations to characterize the  signatures
 of multi-photon-initiated cascades.     The Monte Carlo code ISUSIM 
(Mohanty et al. 1998) was used which  includes the detector model of the Whipple Observatory
 10~m telescope equipped with a $\rm 4.8^{\circ}$ field-of-view 331-photomultiplier camera
(Quinn et al. 1999).   The   underlying goal  was to achieve good  background suppression while 
maintaining maximum detection efficiency.   

A microsecond burst consists of multiple primary  $\gamma$-rays  producing independent 
cascades. 
We  have generated  $\rm 10^8$ individual $\gamma$-rays randomly 
spread out over a range  of impact radius 0 - 1000~m to study the properties
of bursts. The Cherenkov photons which hit the mirror and are reflected  into the
focal plane detector have been superimposed.
In order to trigger on a wavefront event, the number of photoelectrons created in
several  photomultipliers used for  forming a coincidence  has to significantly 
 exceed the number of photoelectrons 
initiated by  fluctuations from the night-sky background.  Therefore, a minimum 
number of $\gamma$-rays per square meter
(Cherenkov photon yield $\rm \sim$ number of primary $\gamma$-rays) is required
to detect a signal in the photomultiplier camera (for trigger specifications, see \S 4).
The fluence required for the detection of a wavefront event of given $\gamma$-ray
energies  is proportional to the number of incoming $\gamma$-rays$\rm /m^{2}$ during 
the time of the burst,  ultimately  determining the number of Cherenkov photons  arriving 
at the detector.

\subsection{Image characteristics}

The technique of recording the Cherenkov images of single-particle-induced showers has proven 
to be effective in distinguishing $\gamma$-ray induced showers from the  more numerous 
background images from cosmic rays.  The usefulness of imaging to identify wavefront events is
addressed in this section. 

Figure~2 shows  the Cherenkov light image from a simulated wavefront event
(300~MeV~$\gamma$-rays traveling parallel to the optic axis)  in the
331-phototube camera of the Whipple Observatory 10~m telescope.  The area  
and the gray-scale of the filled circles indicates the number of photoelectrons  
detected in each pixel.    
The fluence of the event in Figure~2 is $\rm  2.4 \times 10^{-8} ergs/cm^2$ 
 (0.5 $\gamma$-rays$\rm /m^{2}$ at 300 MeV).   
   Figure~3 shows the image of a wavefront event arriving
$\rm 1.13^{\circ}$ off-axis, and it can be seen that the  image is off-set by  
$\approx 1.1^{\circ}$ from the center of the camera.   The image center can be used to measure 
the arrival direction of the burst.  In both cases the light distribution shows a circular image.  
In the case of the image in Figure~3,  the fluence is two times
higher than in Figure~2 and a smoothly decreasing ``halo'' can be seen.  The light beyond the 
central plateau ($\rm 0.3^{\circ}$ in radius) is caused mainly  by the multiple-scattering of 
relatively low energy electrons.  This  halo is not easily recognizable in Figure~2 
(where the burst has a lower fluence),   because the amount of light is comparable to the noise 
fluctuations from the night-sky background.


The structure of the image can be described by its circular shape and its characteristic
radius.    The image shape is described here using a combination of the parameters, 
\textit{Width} and $Length$ (Hillas 1985).  The $Eccentricity$  of an image, characterizing its circular
shape is defined by:   $ Eccentricity = \sqrt{1-Width^2/Length^2}$.  A perfectly circular image
would have an $Eccentricity$ equals zero.  The radial extend of the images  is 
described by $Radius$,  defined  by:  $ Radius = (Width+Length)/2$.   
The  $Radius$  and $Eccentricity$  distribution for wavefront events (individual $\gamma$-rays
 of 200~MeV-5~GeV sampled from a power-law distribution with a differential 
spectral index of -2.5) are shown in Figure~4b and Figure~5a, respectively.  The $Radius$ of the images 
is well defined  and  substantially bigger than for most cosmic-ray showers.
A selection of images with $Radius$ $\rm > 0.70^{\circ}$ would reject most cosmic-ray images. 
The $Eccentricity$ distribution  peaks  at 0.2 (Figure~5a) which corresponds to mostly circular images,  
establishing their circular shape.      
Given  the  $Radius$  and  $Eccentricity$  distribution of recorded cosmic-ray showers 
(dotted line in Figure~4b,~5a),   a strong background suppression can be achieved in the search  
for multi-photon-initiated cascades.


 An additional feature that can be used is the relatively smooth light distribution which is very 
different from most cosmic-ray shower images.
The images from wavefront events reflect the fact that many showers 
 contribute to an image:  their light distribution is extremely
smooth.   The smoothness of an image can be quantified,  e.g., by 
calculating the R.M.S. of the  light content of all pixels.

\subsection{Angular resolution}

 The image center position  provides an estimate of the true arrival 
direction of a wavefront event.  The angle  $\rm \Theta$ is the
difference between the reconstructed and the true arrival direction in degrees. 
Figure 5b  shows the  $\rm \Theta^2$ distribution  of simulated bursts 
with each burst containing energies between 0.2 - 5 GeV sampled from power-law of $\rm E^{-2.5}$.  
The reconstruction accuracy also depends on the total amount of light collected and 
therefore the  fluence of the burst.  The  angular resolution $\rm \sigma_{\Theta}$ is 
defined here so that 72\% of the  image centers would fall within a radius of $\rm \sigma_{\Theta}$. 
  The resolution  for a burst with a fluence of $\rm 1.5 \times 10^{-8} erg/cm^2  $  is 
$\rm \sigma_{\Theta} = 0.12^{\circ}$.  However, for a burst with  a fluence of 
$\rm 6.0 \times 10^{-8} erg/cm^2  $ the resolution is 
 $\rm \sigma_{\Theta} = 0.06^{\circ}$ and improves approximately with the square root of 
the fluence.

\subsection{Timing characteristics}

Images of wavefront events have a characteristic shape,  but even so, image analysis might not
remove the background from  cosmic-ray induced showers completely.   
The pulse shape of the Cherenkov light pulse provides an additional  signature to
identify and distinguish multi-photon-initiated cascades from single-particle-initiated 
air showers.   Pulse shapes from cosmic-ray  air showers are typically a 
few nanoseconds wide. 
Multi-particle-initiated showers from bursts show a minimum time scale of at least 40~ns.

We have used the Whipple Observatory 10~m telescope to record pulse shapes of cosmic-ray 
induced air showers  utilizing a 4-channel digital oscilloscope (Hewlett-Packard~54540A) 
with a 500 Mhz sampling time.  Four channels were used to record pulses from phototubes 
which were spread out over an area of $\rm 0.5^{\circ} \times 0.5^{\circ}$ in the focal plane. 
The trigger requires all four channels to ensure that the system would  only record pulses 
from fairly extended images, similar to multi-photon-initiated events.
   Smaller images can  be distinguished by imaging, 
e.g., by measuring their $Radius$ and $Eccentricity$.      The 
oscilloscope readout was initiated whenever four channels exceeded
a threshold of 30~mV with a time overlap of at least 10~ns.   The length of each
record was chosen to be 2~microseconds with a  time resolution of 4~nanoseconds. 
 The recording system including the photomultiplier, cables and  amplifiers used  
was sensitive to pulse widths ranging from  10~ns up to several hundred~ns.

Figure~6a shows the pulse shape of a typical Cherenkov light flash recorded with 
the Whipple Observatory 10~m telescope.  In comparison we  show (Figure~6b)  the simulated 
pulse profiles from a multi-photon-initiated cascade  from a  100~ns burst of 500~MeV $\gamma$-rays 
of two  different pixels: in the center  of the image (solid line) and a pixel $\rm 1^{\circ}$
 off-center (dashed line).      The pulse profiles of the multi-photon-initiated cascade are
broad and only slightly shifted with respect to each other.  The fluence for the simulated
wavefront event is $\rm 1.1 \times 10^{-7} ergs/cm^2$, about 7 times higher than the 
sensitivity limit of the technique using the Whipple Observatory 10~m telescope. 
It is important to point out that our measurement gives a limit for the background
expected for a burst sensitivity $\rm 1.1 \times 10^{-7} ergs/cm^2$.  Operating at a
lower threshold might imply a higher background.

Figure~7 shows the distribution of pulse widths for a data sample consisting  of 
10,000~events taken during  6~hours of observation time.  The longest pulse
 recorded shows a F.W.H.M. of 33~nanoseconds.    It is important to notice,
that the pulse widths presented here are broadened by 180~foot of
 coax cable (RG-58).  The intrinsic pulse width of Cherenkov pulses are 
somewhat shorter.
This clearly indicates that pulse profiles provide excellent  background discrimination of 
cosmic rays.   Note, that even with this relatively simple set-up a sensitivity of 
$\rm 1.1 \times 10^{-7} ergs/cm^2$ for bursts of  500 MeV $\gamma$-rays would be reached.  

\subsection{Other sources of background}

A  second background showing similar time profiles to those of
multi-photon-initiated cascades could arise from fluorescence light of 
ultra-high energy cosmic-rays (UHECR) at E~$\rm  > 10^{16}$~eV.
Although rare, they could constitute a background of slow pulses.
  The recorded image of the event will help to reject fluorescence events:  
the image would appear as an extended band through the camera, as opposed 
to a circular flat image from a wavefront of multiple $\gamma$-rays.
The pulse profiles of fluorescence light depends  on the impact parameter 
and the arrival direction of the UHECR-shower (Baltrusaitis et al. 1985).  
For the pixellation of the Whipple camera ($\rm 0.25^{\circ}$), the pulse 
width of a UHECR-shower ranges from 70 ns to 350 ns for an impact parameter of 5~km  
and 1~km,   respectively. 
Fluorescence light events can be distinguished from
 wavefront events by  the average arrival  time of photons  (center of the pulse)  in various pixels 
across the field of view.  They  differ according to the geometrical time-of-flight difference 
between the telescope and different parts  of the shower.  
As a consequence, the pulses in different pixels of a fluorescence event 
should be substantially shifted with respect to each other along the shower
axis, whereas the average arrival times of pulses from a wavefront event have a small
intrinsic  time spread and a circular symmetric arrival time pattern.   Therefore, 
it is expected that even with a single telescope,  rare fluorescence events could be 
eliminated.

Light flashes from meteors and lightning have to be considered as a potential source
of background.  The time constant of faint meteors is of order 10 msec or greater
(Cook et al. 1980) and is not in the range of microsecond bursts.   
Lightning pulses are in the range of tens to hundreds of microseconds (Krider 1999).

\section{Trigger criteria}

The properties of images from wavefront events are vastly different than typical
$\gamma$-ray images, for which imaging Cherenkov telescopes are usually optimized.
 Images from TeV $\rm \gamma$-ray primaries exhibit a small angular spread 
 requiring a  trigger sensitive to an elliptical image  extending over an area of
$\rm \approx 0.15^{\circ} \times 0.30^{\circ}$ in the field-of-view.  
Thus Cherenkov telescopes often have a trigger requirement of  a two-fold (four-fold for 
high resolution cameras)  coincidence.
The large angular extent of wavefront  images puts a very different requirement on the trigger
geometry - covering a solid angle of $\rm \approx 1.5^{\circ}$  in diameter.
The limiting factor in both cases is fluctuations from the night-sky background light.
The signal to noise ratio needs to be optimized in order to achieve the highest sensitivity.
In case of wavefront events, the image is bright  within the 
central $\rm 1.5^{\circ}$,  the highest signal-to-noise ratio would be achieved by  
triggering on the total light covering the central $\rm 1.5^{\circ}$ of the image.
 A high-fold coincidence over pixels could be used to trigger efficiently on 
wavefront events  helping  to reduce the random triggers arising from the night-sky background 
light fluctuations.   Using a pixellation  of $\rm  0.25^{\circ}$ (see Figure 2), a  high-fold 
coincidence of 40 pixels would provide a reasonable trigger condition.     

Also, the timing characteristics of the trigger, providing a good sensitivity for wavefront events
is different than for TeV $\gamma$-ray observations.  Single $\gamma$-ray detection  uses a 
typical coincidence time of  10 nanoseconds.   Wavefront event  recording would 
be based on the integration time scale in the order of 100~nanoseconds up to a few microseconds
depending on the putative astrophysical burst time scale.  It is important to point out that in
case of wavefront detection an integration of the signal over the burst time scale
is most efficient to increase the signal-to-noise ratio at the trigger level.
 To search for astrophysical phenomena  whose emission time scale is uncertain a trigger 
operating at several different time
scales in parallel  is necessary, similar to the technique used for the Fly's Eye detector
(Baltrusaitis et al. 1985).

\section{Sensitivity}

The detection of bursts using the imaging technique as described in this paper involves two steps:
triggering on the Cherenkov light flash associated by the multi-photon-initiated cascade 
and discriminating a wavefront event from cosmic-ray showers.   Both requirements impact the 
sensitivity at a given energy and burst time scale.   The sensitivity for a Whipple type 10~m 
telescope equipped with a $\rm 4.8^{\circ}$ field-of-view camera with 331 pixels  is estimated 
(see also Quinn et al. 1999). 

The trigger threshold for the detection of short bursts is a function of the 
fluence of the burst, expressed in $\rm ergs/cm^2$.   The fluence is the product
of the energy of the incoming particles and the number of particles per unit area
impinging on the upper atmosphere.    In order to trigger on a wavefront event 
we require 40 pixels to exceed the night-sky background  fluctuations by  $3 \sigma$.
This not only prevents  triggering  on night-sky background fluctuations, it  also 
ensures a good image reconstruction.      Figure 8 shows the 
fluence sensitivity as a function of energy for 100~ns and 1~$\rm \mu s$  burst 
time scale.
For comparison to previous efforts we also show the sensitivity of the EGRET detector.  EGRET
had a sensitivity for bursts lasting for 600~ns where it  records multiple events 
within one readout cycle.  We have assumed here a collection area of $\rm 0.15 \: m^2$ 
and a minimum of 5 $\gamma$-rays to be detected.  Over the energy range of 300~MeV to
1~GeV the sensitivity of the wavefront technique could exceed EGRET's sensitivity by
a factor of 100 to 500 for 100~ns bursts.

The energy threshold for the detection of wavefront events is limited to lower energies 
by the multiple scattering  angle and by the Cherenkov threshold for radiation by electrons 
of 90~MeV at 20~km atmospheric height.   This results in a natural barrier for the atmospheric 
Cherenkov technique.  We have limited our simulations  to energies between 
200~MeV  to 5~GeV.

\section{Summary}

We have shown that  sub-GeV $\gamma$-ray bursts  lasting for $\rm >$~100 nanoseconds 
to microseconds could be efficiently detected using a single ground-based  imaging 
Cherenkov telescope.   
The technique described is based on previous attempts to detect multi-photon-initiated 
cascades  from short bursts.  
However,  we show for the first time that the angular Cherenkov light distribution together 
with the  pulse shape can be used to advantage to search for short bursts with a single 
imaging telescope.

Measurements of Cherenkov pulse shapes of cosmic-ray induced showers indicates that 
pulse shapes from multi-photon-initiated cascades are well separated from background showers. 
  A search for microsecond bursts would use this criterion  as a first filter.  If events with
 long pulse durations were found, image analysis could verify if those events are consistent 
with the very distinct image shapes of a multi-photon-initiated cascade.  The image also 
contains valuable information of the arrival direction with an angular resolution of 
$\rm 0.06^{\circ}-0.12^{\circ}$, depending on the fluence.
The fluence sensitivity of the Whipple telescope with a microsecond trigger exceeds 
EGRET's sensitivity by more than two orders of magnitude.

In addition, arrays of telescope could be used to further improve this technique. 
In contrast to air showers, wavefront events would appear identical in the field-of-view
of  arrays of telescopes with a typical spacing of $\rm \approx 100$~m.   Single-particle initiated
air showers show a  parallactic displacement because of the different distances to the
shower core.   
In view of several proposed next generation detectors (VERITAS, HESS; overview see 
Krennrich 1999b), the implementation of this technique in  telescope arrays could provide 
the highest  fluence sensitivity of any existing $\gamma$-ray detector for microsecond  bursts at 
sub-GeV - several-GeV energies.

\vfill\eject

\acknowledgments
 
This research is supported by grants from the U.S. Department of Energy.

\vfill\eject

{}

\clearpage

\figcaption[m42d]
{(a)~The longitudinal and lateral distribution of the electromagnetic component of 
a  single $\gamma$-ray initiated shower of 1 TeV,  traced by the
Cherenkov light is shown.  The dots indicate the origin of emission of individual
Cherenkov photons that are detected with a Whipple type telescope, located at an
elevation of 2306~m and at X-coordinate zero.    
(b)~We show  the corresponding  distribution for a multi-$\gamma$-ray-initiated 
shower (note that lateral scale is a factor of 10 larger).  
The single 1~TeV $\gamma$-ray produces a narrow band Cherenkov photon distribution in the 
atmosphere.  It  can  be   detected up to a distance of 150~m from the shower core.
For the  multi-$\gamma$-ray-initiated shower,  Cherenkov photons that originate up to 600~m away
in the lateral scale can still contribute to the Cherenkov flash detected in a  telescope.  
}

\figcaption[m42d]
{The simulated image of a burst of 300 MeV $\gamma$-rays ($\rm 2.4 \: \times 10^{-8} erg/cm^2$)  arriving  within
100~nanoseconds as it would be seen with the Whipple Observatory 10~m telescope.  The area and the gray-scale of the
filled circles indicate the relative light content in each pixel (maximum of 35 photoelectons per pixel in this event).  Night-sky noise fluctuations
for a 100~ns integration time are included.  The image has been processed using the standard 
image cleaning procedure (Reynolds et al. 1993).  The circle indicates the angular extension and the shape 
 of the image.  The center of the circle coincides with the arrival direction of the burst 
to within $\rm 0.1^{\circ}$.}

\figcaption[m42d]
{The simulated image of a burst of 300 MeV $\gamma$-rays lasting for
100~ns with a photon density of (fluence = $\rm 4.8 \: \times 10^{-8} erg/cm^2$).   
The  arrival direction was offset by $\rm 1.13^{\circ}$ from the optic axis of the
telescope.   Image  processing  using the standard image cleaning procedure (Reynolds et al. 1993)
has been applied.  The circle indicates the plateau and the drop in the light density
 of the image where its center coincides with the arrival direction of the burst.  
The smooth ``halo'' surrounding the central image is the other
characteristic feature of a burst when its light content is significantly above the night sky noise.  }

\figcaption[m42d]
{ (a) The average radial light profile (light density vs. radial distance from image center) 
of wavefront events is shown.  (b) The estimated $Radius$ of 
simulated wavefront events (solid line) is compared with the $Radius$ of detected cosmic-ray
background events (dashed line).  Only cosmic-ray  events with the same or larger light
content (size) in the image as for the simulated wavefront events  are accepted.
The average $Radius$ of the images from 500~MeV bursts is approximately $\rm 0.8^{\circ}$ which corresponds 
to the half width in the radial profile. }

\figcaption[m42d]
{(a) The $Eccentricity$ ($ \sqrt{1-Width^2/Length^2}$) of  images from wavefront events are shown.  
The distribution for wavefront events  peaks at 0.2 as expected for almost circular 
images.  The dotted curve represents cosmic-ray showers recorded with the Whipple Observatory
10~m telescope.
(b) The square of the difference between the  reconstructed arrival direction from the true 
arrival direction ($\rm \Theta^2$) of wavefront events is plotted.   The angular resolution 
is  $\rm 0.12^{\circ}$ for a burst  close to the detection threshold (solid line) and becomes
$\rm 0.06^{\circ}$  for bursts with 4 times higher fluence values (dotted line).  Background 
images from  cosmic rays with isotropic arrival directions  would show a flat distribution 
in this representation.  }

\figcaption[m42d]
{(a) The pulse shape for a cosmic-ray  event recorded with the Whipple 10~m telescope is shown.
The noise is due to fluctuations from the night sky background light.
  (b)~The pulse profile of a simulated  multi-photon-initiated cascade for 2 photomultipliers
     one in the center of the image (solid line) and one by $\rm 1^{\circ}$ off-center (dotted line)
     are shown.    The burst time scale is 100~ns.   Here the night sky background  noise is not included,
 but it would be comparable to the noise in figure 5a.  }

\figcaption[m42d]
{The pulse width distribution for events recorded with the Whipple 10~m telescope.
  The expected range for bursts from  primordial black holes would be above 100 ns.
  The shortest possible pulse width from a multi-photon-initiated cascade that occurs 
    due to arrival time differences between the sub-showers is at about 40 ns. }

\figcaption[m42d]
{The fluence sensitivity for the wavefront technique is shown as a function of energy.  For 
comparison we also show the fluence sensitivity for EGRET,  with a collection
area of $\rm 0.15 \: m^2$ in the given energy range.  The detection of at least 5 $\gamma$-rays
has been required.  It can be seen that the wavefront technique is 
about two orders of magnitude more sensitive than EGRET, which is mainly limited by its collection 
area. }

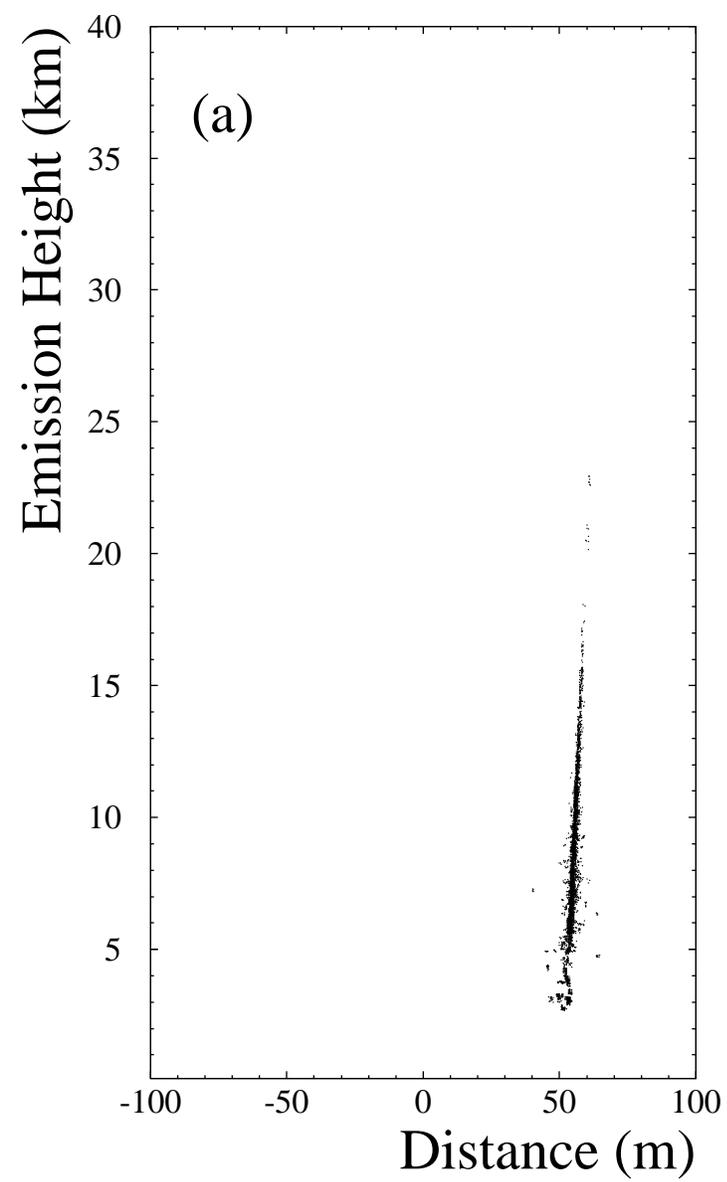
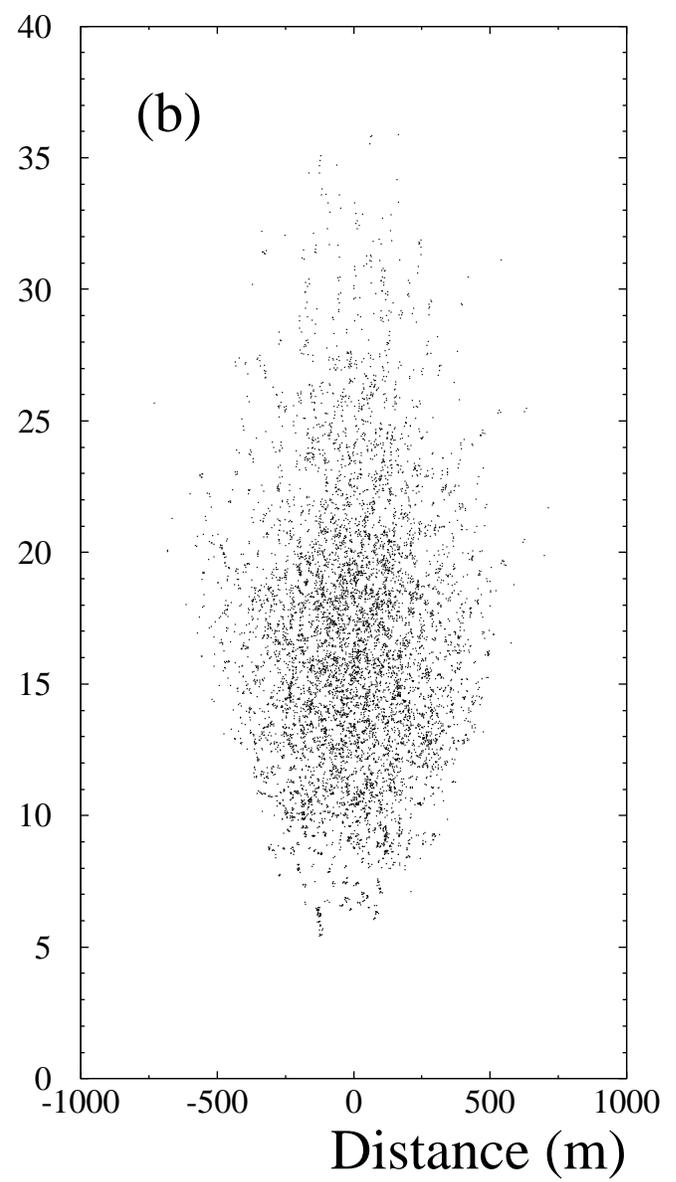
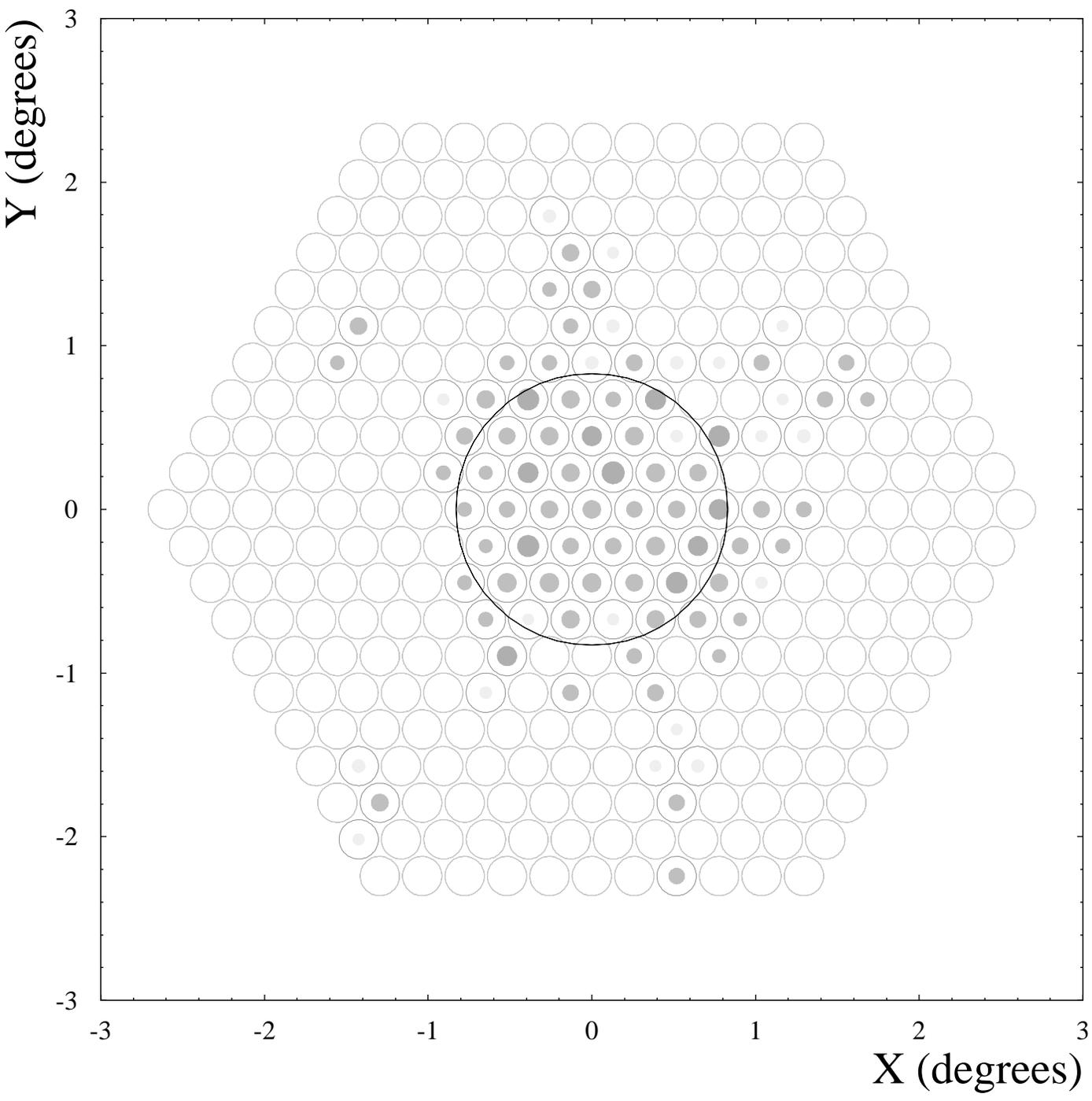
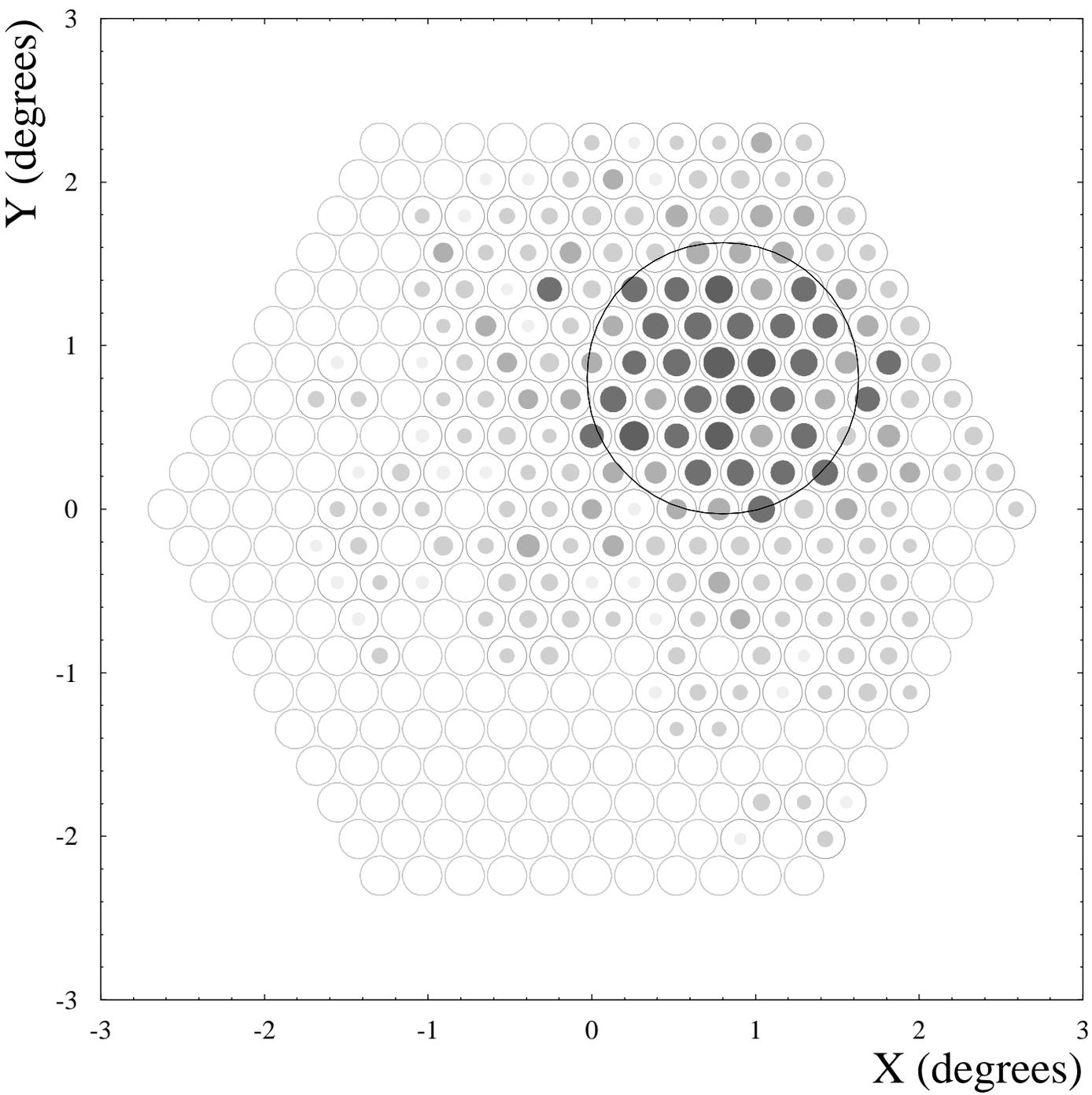
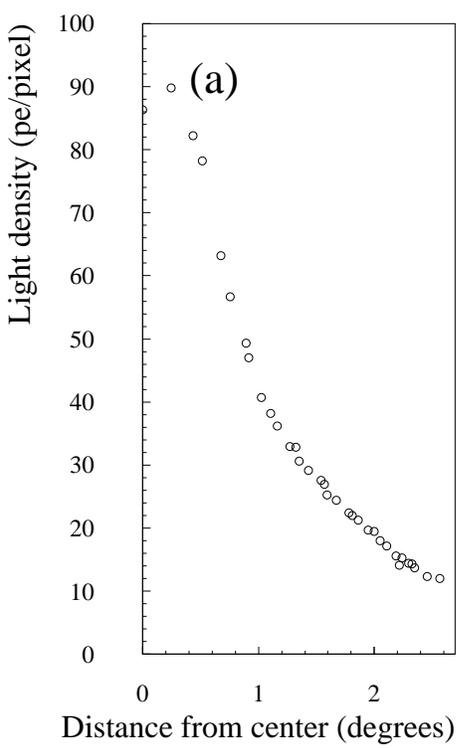
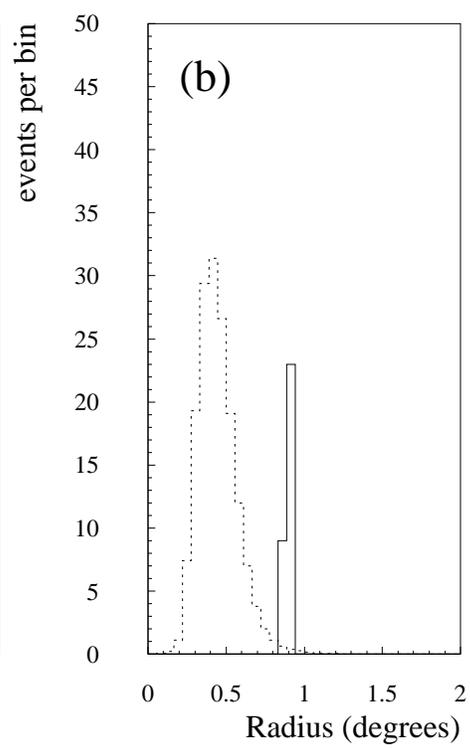
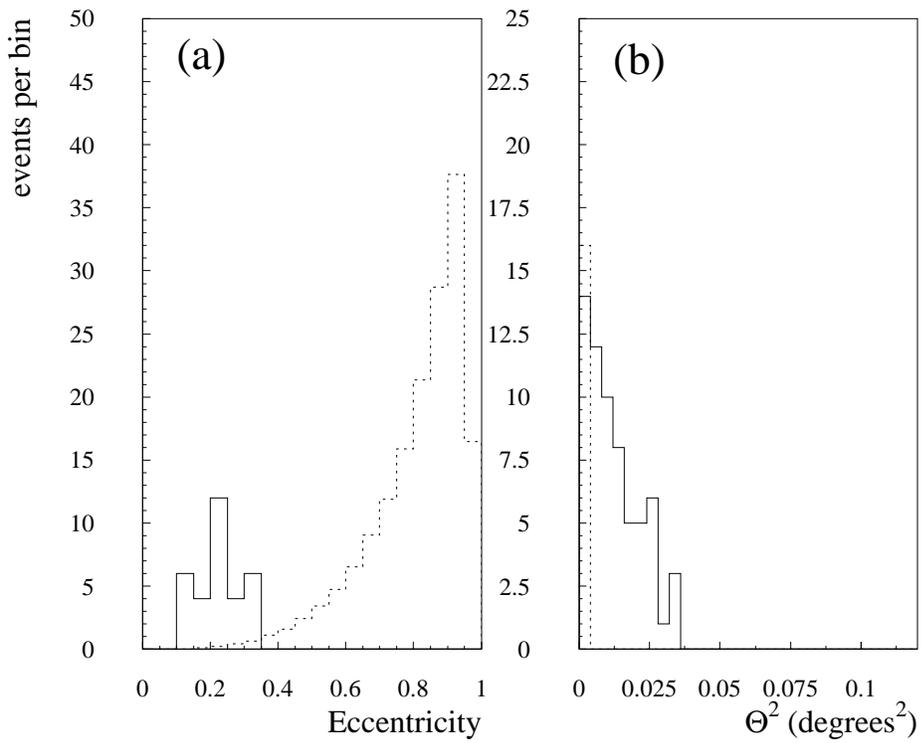
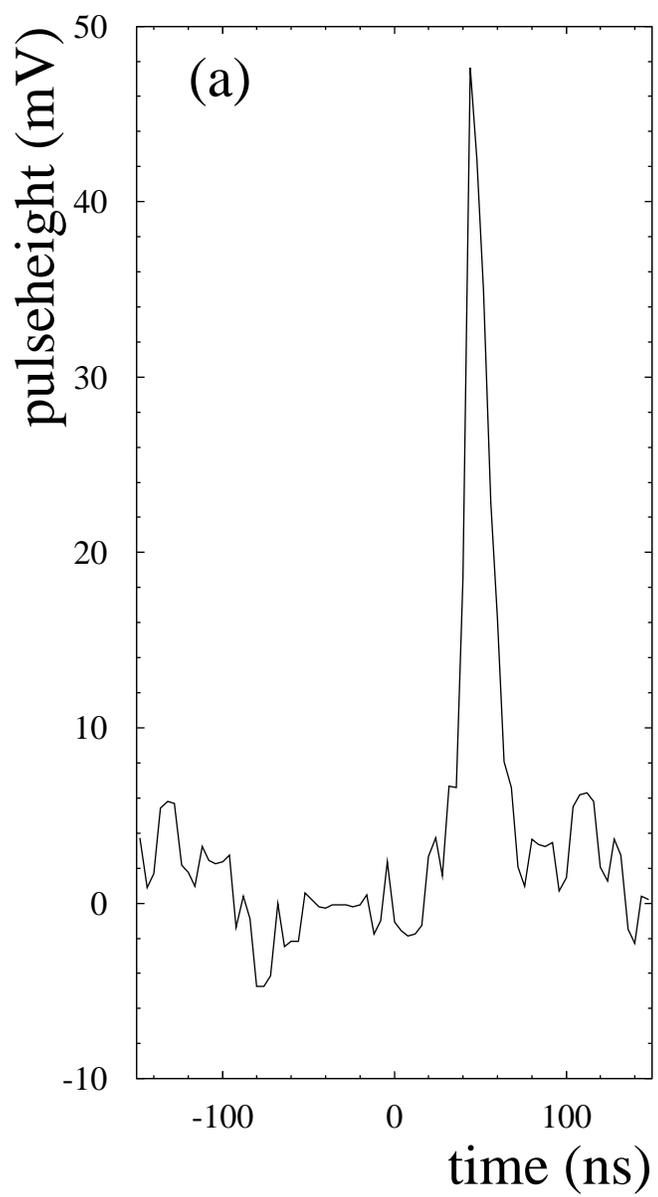
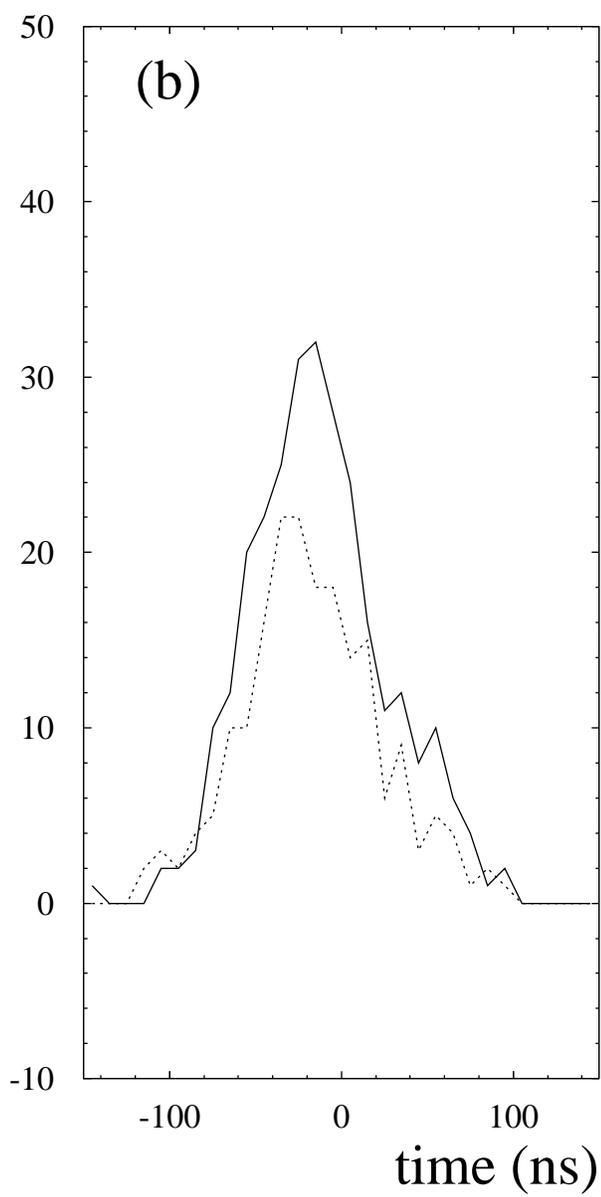
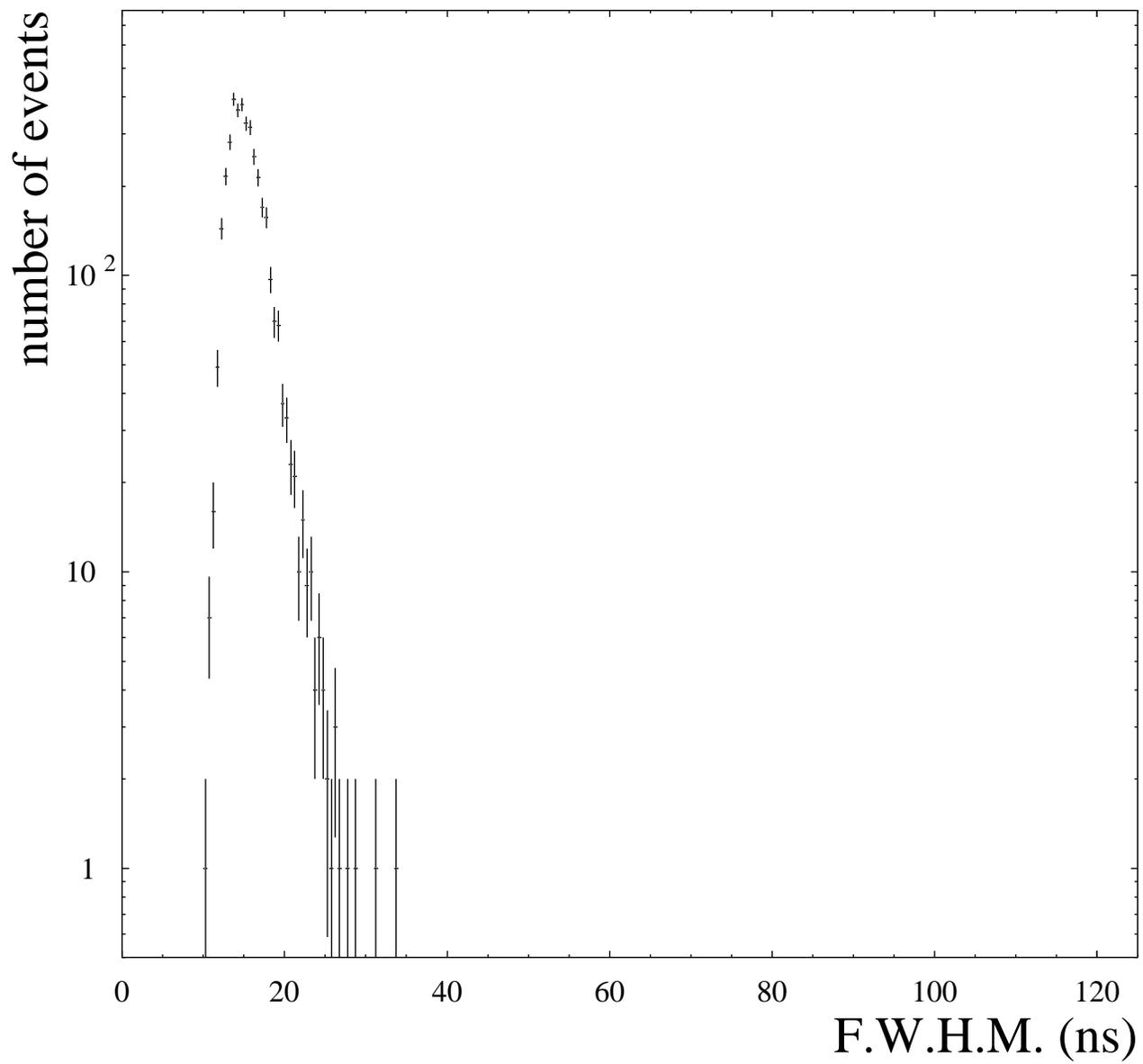
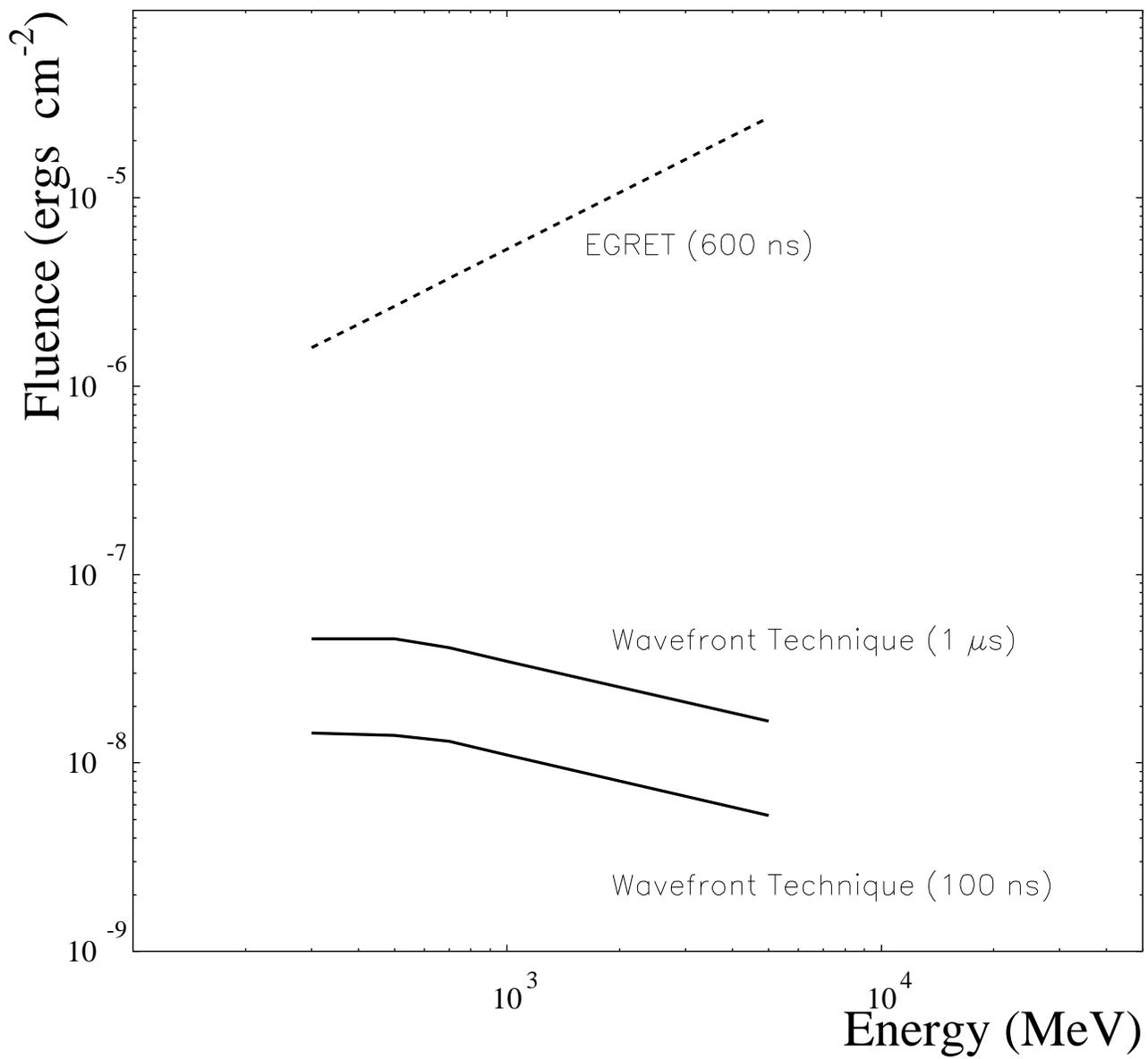
\end{document}